\documentclass[12pt]{article}
\usepackage[dvips]{graphicx}
\usepackage{amssymb}

\makeatletter
 \@addtoreset{equation}{section}
 
\makeatother

\newcommand{\maru}{\mbox{\tiny$\stackrel{\circ}{\scriptstyle\circ}$}}

\begin{document}


\begin{titlepage}

\renewcommand{\thefootnote}{\fnsymbol{footnote}}

\begin{flushright}
\begin{tabular}{l}
UTHEP-563\\
RIKEN-TH-130\\
May 2008
\end{tabular}
\end{flushright}

\bigskip

\begin{center}
{\Large \bf
D-brane States and Annulus Amplitudes in

$OSp$ Invariant Closed String Field Theory
}
\end{center}

\bigskip

\begin{center}
{\large
Yutaka Baba}${}^{a}$\footnote{e-mail:
        ybaba@riken.jp},
{\large Nobuyuki Ishibashi}${}^{b}$\footnote{e-mail:
        ishibash@het.ph.tsukuba.ac.jp},
{\large Koichi Murakami}${}^{a}$\footnote{e-mail:
        murakami@riken.jp}
\end{center}

\begin{center}
${}^{a}${\it Theoretical Physics Laboratory, RIKEN,\\
         Wako, Saitama 351-0198, Japan}
\end{center}
\begin{center}
${}^{b}${\it
Institute of Physics, University of Tsukuba,\\
Tsukuba, Ibaraki 305-8571, Japan}\\
\end{center}

\bigskip

\bigskip

\bigskip

\begin{abstract}
In the $OSp$ invariant closed string field theory,
we construct the states corresponding to 
parallel D-branes that are located at different points
in the space-time.
Using these states, we evaluate annulus amplitudes.
We show that the results coincide with those of
first quantized string theory.

\end{abstract}

\setcounter{footnote}{0}
\renewcommand{\thefootnote}{\arabic{footnote}}

\end{titlepage}

\section{Introduction}

D-branes have been playing an important role in understanding
nonperturbative aspects of string theory.
In previous works \cite{Baba:2006rs}\cite{Baba:2007tc},
we studied how to describe D-branes
in closed string field theory.
The closed string field theory that we consider is
the $OSp$ invariant string field theory
for bosonic strings \cite{Siegel:1984ap}.
(See also \cite{Neveu:1986cu}\cite{Uehara:1987qz}
\cite{Kugo:1987rq}\cite{Kawano:1992dp}.)
We constructed
the states with an arbitrary number of coincident
D-branes and ghost D-branes \cite{Okuda:2006fb}
in this closed string field theory.
We can calculate disk amplitudes using these states, and
the results coincide with those of first quantized string theory~\cite{Baba:2007tc}.

In this paper, 
we extend our construction into the case
where the D-branes are located at different points from each other
in the space-time.
Using such a state with two D-branes, we evaluate 
annulus amplitudes.
We show that they coincide with the usual annulus amplitudes
including the normalizations. 
This fact yields another evidence for our construction. 

The organization of this paper is as follows.
In section \ref{sec:ND-braneState}, we generalize our
previous construction \cite{Baba:2007tc} to propose
the states for $N$ parallel D$p$-branes
that are located at different points from each other.
We show that these states are BRST invariant in the leading order
in the regularization parameter $\epsilon$.
In section \ref{sec:annulusamp},
we compute annulus amplitudes
and show that the results in first quantized string theory are reproduced.
Section \ref{sec:discussions} is devoted to conclusions.
In appendix \ref{sec:detailcalculation},
we present details of the calculation.

\section{States with Parallel D-branes at Different Points}
\label{sec:ND-braneState}
The BRST invariant state corresponding to one flat D$p$-brane
sitting at $X^{i}=Y^{i}$ $(i=p+1,\ldots,25)$
is constructed in \cite{Baba:2007tc}\footnote{In this paper, 
the notations for the $OSp$ invariant
string field theory are the same as those used in
\cite{Baba:2007tc}, unless otherwise stated.} 
as
\begin{eqnarray}
|D_{+} (Y)\rangle \! \rangle
 = \lambda 
  \left( \int d\zeta
         \bar{\mathcal{O}}_{D}
         \left( \zeta , Y \right)
  \right)
  |0\rangle \! \rangle~,
\label{eq:1Dbrane}
\end{eqnarray}
where
\begin{eqnarray}
&& \bar{\mathcal{O}}_{D} (\zeta, Y )
    = \exp \left[ A \int^{0}_{-\infty} dr\,
       \frac{e^{\zeta \alpha_{r}}}{\alpha_{r}}
       \,
       {}^{\epsilon}_{r} \langle B_{0}(Y) |
          \bar{\psi}\rangle_{r} + B\zeta^{2} \right]~,
\nonumber\\
&& \qquad
 A=\frac{(2\pi)^{13}}{(8\pi^{2})^{\frac{p+1}{2}} \sqrt{\pi}}~,
\quad
 B=\frac{(2\pi)^{13}\epsilon^{2} (-\ln \epsilon)^{\frac{p+1}{2}}}
        {16 \left(\frac{\pi}{2}\right)^{\frac{p+1}{2}}
         \sqrt{\pi}g}~.
\end{eqnarray}
Here 
$|B_{0}(Y) \rangle \equiv e^{-ip_{i} Y^{i}} |B_{0} \rangle$
denotes the boundary state for
the D$p$-brane located at $X^{i}=Y^{i}$
and  $|B_{0}\rangle = |B_{0} (0) \rangle$
is given in \cite{Baba:2007tc}.
As in \cite{Baba:2007tc}, we introduce the state
\begin{equation}
|B_{0}(Y)\rangle^{T}
 = e^{-\frac{T}{| \alpha |} (L_{0} + \tilde{L}_{0} -2)}
  |B_{0} (Y) \rangle~,
\end{equation}
and use $|B_{0}(Y)\rangle^{\epsilon}$ with
$0<\epsilon\ll 1$ as a regularized version of
$|B_{0}(Y)\rangle$.
$\int d\zeta \bar{\mathcal{O}}_{D}$ can be considered as 
an operator which creates the D-brane by acting on the 
second quantized vacuum $|0\rangle \! \rangle$. 
With string field $|\bar{\psi}\rangle$ exponentiated, 
this operator has the effect of inserting boundaries in 
the worldsheet.

We would like to show that the states corresponding to 
$N$ such D$p$-branes located at 
$X^{i}=Y^{i}_{(I)}$ $(I=1,\ldots,N)$
can be given simply as 
\begin{eqnarray}
|D_{N+};Y_{(I)}\rangle \! \rangle
 = \lambda_{N+} \prod_{I=1}^{N}
  \left( \int d\zeta_{I}
         \bar{\mathcal{O}}_{D}
         \left( \zeta_{I},Y_{(I)} \right)
  \right)
  |0\rangle \! \rangle~,
\label{eq:NDbranes}
\end{eqnarray}
if 
$(Y^{i}_{(I)} - Y^{i}_{(J)})^{2} \neq 0$ for $I\neq J$. 
In contrast to the case of coincident D-branes studied
in \cite{Baba:2007tc}, 
we just have to consider the product of 
$\int d\zeta \bar{\mathcal{O}}_{D} $. 
Indeed, we can show that as long as
$(Y^{i}_{(I)} - Y^{i}_{(J)})^{2} \neq 0$ for $I\neq J$,
the states (\ref{eq:NDbranes}) are BRST invariant
in the leading order in the regularization parameter $\epsilon$.
The proof goes exactly as in \cite{Baba:2007tc}. 
One crucial difference from the coincident case 
is that in the limit of $T=\epsilon \rightarrow 0$
the string vertex
\begin{equation}
\langle V_{1}(3);Y,Y';T|
 \equiv \int d'1 d'2 \, \langle V_{3}(1,2,3)|
     B_{0}(Y)\rangle^{T}_{1}
     |B_{0}(Y')\rangle^{T}_{2}
\label{eq:V1YYT-1}
\end{equation}
is suppressed by
$\epsilon^{\frac{(\Delta Y^{i})^{2}}{4\pi^{2}}}$
with $\Delta Y^{i} \equiv Y^{i} - Y^{\prime i}$,
compared with
$\langle V_{1}(3);T|$ evaluated in \cite{Baba:2007tc}.
Because of this suppression, 
the interaction between $\bar{\mathcal{O}}_{D}$ at different 
points can be ignored in the leading order in $\epsilon$ and 
the states (\ref{eq:NDbranes}) become
BRST invariant. 

The details of the calculation of 
$\langle V_{1}(3);Y,Y';T|$ are presented in 
appendix \ref{sec:detailcalculation}. 
The suppression stated above is intuitively obvious, 
because the D-branes sit at different points. 
The suppression factor originates from 
the factor $e^{-S_{\mathrm{cl}}}$, where $S_{\mathrm{cl}}$ is
the classical action given in eq.(\ref{eq:classicalaction})
on the worldsheet depicted in Fig.~\ref{fig:worldsheet}
in appendix~\ref{sec:detailcalculation}.
Indeed, using the results in
\cite{Kishimoto:2004km}\cite{Kishimoto:2006gd}\cite{Baba:2007tc},
we find that in the $T=\epsilon \rightarrow 0$ limit
\begin{equation}
e^{-S_{\mathrm{cl}}}
 \sim \left( \frac{\epsilon}
                  {4\alpha \sin (-2\pi V_{3})}
      \right)^{\frac{(\Delta Y^{i})^{2}}{4\pi^{2}}}~.
\label{eq:sclepsilon}
\end{equation}

In addition to the BRST invariance mentioned above,
it is easy to see that
using the states (\ref{eq:NDbranes})
one can calculate the disk amplitudes
in the same way as in \cite{Baba:2007tc} and obtain 
those for the parallel D-branes. 
Thus we may regard the states (\ref{eq:NDbranes}) 
as the ones where such D-branes exist. 
We can also generalize the states (\ref{eq:NDbranes})
to include ghost D-branes \cite{Okuda:2006fb}, 
as is carried out in \cite{Baba:2007tc}.

\section{Annulus Amplitudes Derived from D-brane States}
\label{sec:annulusamp}

\subsection{Amplitudes with one closed string external line}

\begin{figure}[htb]
\begin{center}
\includegraphics[height=14em]{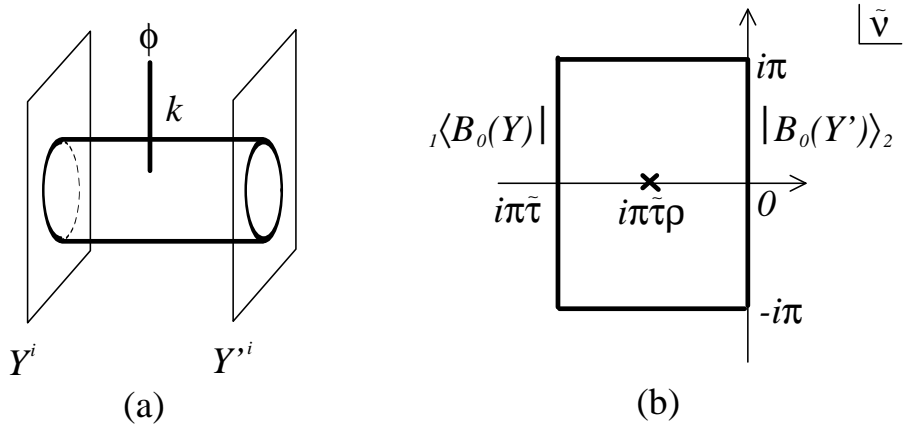}
\end{center}
\caption{(a) The annulus diagram with one closed string
             external state $\phi$.
         (b) The $\tilde{\nu}$ coordinate on the worldsheet of the
             string diagram in (a).
         At $\tilde{\nu}=i\pi\tilde{\tau}\varrho$, the vertex operator
         $V_{\phi}$ corresponding to the state $\phi$ is inserted.}
\label{fig:annulus}
\end{figure}

Using the states with D-branes constructed in the last section,
we would like to calculate scattering amplitudes involving 
the strings whose worldsheets have boundaries attached to D-branes 
contained in these states.
In this paper, we evaluate annulus amplitudes.
Let us first consider the annulus amplitudes with 
one closed string external line as described in 
Fig.~\ref{fig:annulus} (a),
in the situation where the annulus is suspended between 
two parallel D$p$-branes located at $X^{i}=Y^{i}$ and 
$Y^{\prime i}$.
The S-matrix element for this process can be obtained from  
the following correlation function involving these two D$p$-branes:
\begin{equation}
\left\langle \! \left\langle \mathcal{O}_\phi (t,k)
\right\rangle \! \right\rangle_{D_{2+}(Y;Y')}
\equiv
\frac{\langle \! \langle 0 |
        \mathcal{O}_\phi (t,k) |
      D_{2+};Y,Y' \rangle \! \rangle}
     {\langle \! \langle 0 |
       D_{2+};Y,Y'   \rangle \! \rangle}~,
\label{eq:onetachyon}
\end{equation}
where $t > 0$.
$\mathcal{O}_\phi (t,k)$ is the observable corresponding
to the external state $\phi$ defined \cite{Baba:2007je} as
\begin{equation}
\mathcal{O}_{\phi } (t,k) = \int dr \, \frac{1}{\alpha_{r}}
  \, {}_{r}\Big(
        {}_{C,\bar{C}}\langle 0| 
          \otimes {}_X \langle \mbox{primary}_\phi;k | \Big)
            |  \Phi (t)\rangle_{r}~,
\label{eq:tachyon}
\end{equation}
where $|\mbox{primary}_\phi;k \rangle_X$ 
is a normalized Virasoro primary state with momentum $k$, 
corresponding to a particle with mass $M$. 
In the correlation function (\ref{eq:onetachyon})
we should evaluate the contribution
$G_{\phi DD'}(k)$  from the annulus diagram
suspended between the two D$p$-branes contained in
$|D_{2+};Y,Y'\rangle\!\rangle$.
This is an order $\mathcal{O}(g)$ term in the correlation
function (\ref{eq:onetachyon}).

Perturbatively, for the $\zeta_{I}$ $(I=1,2)$ integrations in
eq.(\ref{eq:NDbranes}) the saddle point method becomes a
good approximation \cite{Baba:2006rs}\cite{Baba:2007tc} and yields
\begin{equation}
\left| \left. D_{2+};Y,Y'
      \right\rangle \! \right\rangle
\simeq
 \lambda' \exp \left[
 A \int^{0}_{-\infty} \frac{dr_{1}}{\alpha_{r_{1}}}
  \, {}_{r_{1}}^{\epsilon} \langle B_{0}(Y)|
    \bar{\psi} \rangle_{r_{1}}
   \right]
 \exp \left[
  A \int^{0}_{-\infty} \frac{dr_{2}}{\alpha_{r_{2}}}
  \, {}_{r_{2}}^{\epsilon} \langle B_{0} (Y')|
    \bar{\psi}  \rangle_{r_{2}}
  \right]
  |0\rangle \! \rangle~,
\end{equation}
where $\lambda' = -\frac{\pi}{B} \lambda_{2+}$.
Using these, we obtain
\begin{eqnarray}
\lefteqn{
G_{\phi DD'}(k)
 = - 3! A^{2} 
   \int^{\infty}_{0} \frac{d1}{\alpha_{1}}
   \int^{\infty}_{0} \frac{d2}{\alpha_{2}}
   \int^{0}_{-\infty} \frac{d3}{\alpha_{3}}
  \int^{t}_{0} dT \, \frac{2g}{3}
    \langle V_{3}^{0} (1,2,3)|
}
\nonumber\\
&& 
 \hspace{5em}
 \times
     |B_{0} (Y)\rangle_{1}^{T}
     |B_{0} (Y') \rangle_{2}^{T}
      \, e^{ \frac{t-T}{\alpha_{3}}
         (L^{(3)}_{0} + \tilde{L}^{(3)}_{0} -2)}
      \Big( |\mbox{primary}_\phi ;k\rangle_X \otimes 
       |0\rangle_{C,\bar{C} } \Big)_3~,
\label{eq:GphiDD}
\end{eqnarray}
where $T$ corresponds to the proper time of the 
three-string interaction vertex.
We have performed the Wick rotation so as to make the proper time
Euclidean. 
Using eqs.(\ref{eq:V10YYT}) and (\ref{eq:LPPrectangle}),
the right hand side of eq.(\ref{eq:GphiDD}) can be rewritten as
\begin{eqnarray}
\lefteqn{G_{\phi DD'}(k) 
= -g  A^{2} 
  \int_{0}^{\infty} d\alpha  
  \int^{\alpha}_{0} d\alpha_{1}
  \int^{t}_{0} dT 
  \int \frac{d^{26}p_3}{(2\pi )^{26}}
  id\bar{\pi}_0^{(3)}d\pi_0^{(3)}
\,
\mathcal{K}_1(3;Y,Y';T)
}
 \nonumber\\
&&  \times  
(2\pi)^{p+1} \delta_{\mathrm{N}}^{p+1}(p_3) 
 \langle V^{0}_{1,\mathrm{LPP}} (3); Y,Y';T|
   e^{-\frac{t -T }{\alpha} (L^{(3)}_{0} + \tilde{L}^{(3)}_{0} -2)}
   \Bigl( |\mbox{primary}_\phi ;k\rangle_X \otimes 
       |0\rangle_{C,\bar{C} } \Bigr)_3 ,~~~
\label{eq:Gtdd-3p}
\end{eqnarray}
where  $\alpha = - \alpha_{3}$,
$\mathcal{K}_1(3;Y,Y';T)$ is a factor given 
in eq.(\ref{eq:partition-annulus}) and 
$\delta^{p+1}_{\mathrm{N}} (p_3)$ denotes the delta function
of the momentum in the directions along the D$p$-branes.
In the following, 
we would like to rewrite the right hand side of 
eq.(\ref{eq:Gtdd-3p}) into a form which 
can be compared with the usual annulus amplitude.

\subsubsection*{LPP vertex
               $\langle V^{0}_{1,\mathrm{LPP}}(3) ; Y,Y';T|$}

$\langle V^{0}_{1,\mathrm{LPP}}(3) ; Y,Y';T|$ can be expressed as
a direct product of a state in the $C,\bar{C}$ Fock space and 
one in the $X$ Fock space,
namely
\begin{equation}
{}_{C,\bar{C}}\langle V^{0}_{1,\mathrm{LPP}}(3) ; Y,Y';T|
\otimes
{}_{X} \langle V^{0}_{1,\mathrm{LPP}}(3) ; Y,Y';T|~.
\end{equation}
Since
${}_{C,\bar{C} } \langle V^{0}_{1,\mathrm{LPP}}(3) ; Y,Y';T|$
has the form 
\begin{equation}
 {}_{C,\bar{C}}\langle V^{0}_{1,\mathrm{LPP}}(3) ; Y,Y';T|
 = {}_{C,\bar{C} } \langle 0| 
        e^{- \frac{T}{\alpha} 
               2i \pi_0^{(3)}\bar{\pi}_0^{(3)} 
                     + 
               \mbox{\footnotesize{ (terms quadratic 
                            or linear in oscillators)} 
             }}~,
\end{equation}
the contribution from the $C,\bar{C}$ sector to
$G_{\phi DD'}(k)$ in eq.(\ref{eq:Gtdd-3p}) becomes
\begin{equation}
\int id\bar{\pi}_0^{(3)}d\pi_0^{(3)}
{}_{C,\bar{C} } \langle V^{0}_{1,\mathrm{LPP}}(3) ; Y,Y';T|
0\rangle_{C,\bar{C}}
\, e^{- \frac{t-T}{\alpha}2i\pi_0^{(3)}\bar{\pi}_0^{(3)}}
=
\frac{2t}{\alpha}~.
\end{equation}

{}From the definition of the LPP vertex \cite{LeClair:1988sp},
one can see that the overlap
\begin{equation}
\int \frac{d^{26}p_3}{(2\pi )^{26}}(2\pi)^{p+1}
 \delta_{\mathrm{N}}^{p+1}(p_3)
{}_{X} \langle V^{0}_{1,\mathrm{LPP}}(3) ; Y,Y';T
      | \mbox{primary}_\phi ;k\rangle_{X,3}~
\label{eq:XVLPP}
\end{equation}
is written in terms of a correlation function on the annulus. 
In order to express the correlation function 
using the boundary states, it is convenient to use 
the worldsheet coordinate $\tilde{\nu}$ depicted in
Fig.~\ref{fig:annulus} (b),
which is related to the coordinate $\nu$ in
Fig.~\ref{fig:worldsheet} (b) by
\begin{equation}
  \tilde{\nu} = \frac{2\pi }{- i \tau} \nu~.
\label{eq:tilde-nu}
\end{equation}
In this coordinate, the annulus diagram in
Fig.~\ref{fig:annulus} (a)
is described as a cylinder of circumference $2\pi$.
The length of the cylinder is $-i\pi \tilde{\tau}$ and
the vertex operator corresponding to the external state 
is inserted at 
\begin{equation}
\tilde{\nu} = \frac{2\pi }{- i \tau} V_3
            = i\pi\tilde{\tau}\varrho~,
\end{equation}
where
\begin{equation}
\tilde{\tau}
= 
- \frac{1}{\tau}~, \qquad 
\varrho 
= - 2V_{3} = \frac{\alpha_1}{\alpha}~.
\label{eq:varrho}
\end{equation}
The overlap (\ref{eq:XVLPP}) can be written as
a correlation function\footnote{In the expression 
 ${}_X\langle B_0(Y^\prime )|
  e^{i\pi \tilde{\tau}\varrho
     \left(L_0^X+\tilde{L}_0^X-2\right)}
  \, V_\phi \,
  e^{i\pi \tilde{\tau} (1-\varrho) 
      \left(L_0^X+\tilde{L}_0^X-2\right)}
  |B_{0}(Y)\rangle_X$, 
the integrations over the zero modes $p$ are included 
in the definition of the inner product, 
as is usual in CFT. 
} 
on the cylinder with the coordinate $\tilde{\nu}$
as follows:
\begin{eqnarray}
\lefteqn{
\int \frac{d^{26}p_3}{(2\pi )^{26}}
(2\pi)^{p+1} \delta_{\mathrm{N}}^{p+1}(p_3)
{}_{X} \langle V^{0}_{1,\mathrm{LPP}}(3) ; Y,Y';T
       |\mbox{primary}_\phi ;k\rangle_{X,3}
\nonumber
}\\
&&\quad  =
N(\tilde{\tau}) \left|
\frac{\partial w_3}{\partial \tilde{\nu}}
(i\pi\tilde{\tau}\varrho )\right|^{-\left(k^2+M^2+2\right)}
\nonumber\\
&& \quad \qquad \times
{}_X\langle B_0(Y^\prime )|
e^{i\pi \tilde{\tau} \varrho
\left(L_0^X+\tilde{L}_0^X-2\right)}
\, V_\phi \,
e^{i\pi \tilde{\tau} (1-\varrho) 
\left(L_0^X+\tilde{L}_0^X-2\right)}
|B_{0}(Y)\rangle_X~.
\label{eq:BVB}
\end{eqnarray}
Here $L^{X}_{0}$ and $\tilde{L}_{0}^{X}$ are the zero-modes
of the Virasoro generators and $|B_{0}(Y)\rangle_{X}$ is
the boundary state in the $X$ sector given in \cite{Baba:2007tc}.
$V_\phi$ denotes the vertex operator of weight
$(\frac{1}{2}(k^{2}+M^{2}+2),\frac{1}{2}(k^{2}+M^{2}+2))$
corresponding to the state 
$|\mbox{primary}_\phi ;k\rangle_X $. 
$N(\tilde{\tau})$ is a normalization factor independent of $\phi$, 
which can be fixed by considering the case $V_\phi =1$, 
and we obtain
\begin{equation}
N(\tilde{\tau}) = \eta (\tilde{\tau})^{24}
\prod_{n=1}^\infty 
\left(
1-e^{2\pi i\tilde{\tau}n}
\right)^2
e^{S_{\mathrm{cl}}}(2\pi )^{26-(p+1)}
(-i\tilde{\tau} )^{13-\frac{p+1}{2}}~.
\end{equation}
By using eqs.(\ref{eq:tilde-nu}), (\ref{eq:unitdisk}) 
and (\ref{eq:Mandelstam-annulus}),
we also obtain
\begin{equation}
    \frac{\partial w_3}{\partial \tilde{\nu} }
      (i\pi \tilde{\tau} \varrho)
  =  \frac{-i\tau}{2\pi }
     \left. \frac{\partial w_3}{\partial \nu }
   \right|_{ \nu  =V_3 }
  =
  -i \tau \frac{\eta(\tau)^3}{\vartheta_1 (2V_3|\tau)}
      e^{\frac{T}{\alpha}}~.
\end{equation}

\subsubsection*{Integration measure}
In eq.(\ref{eq:Gtdd-3p}),
we should change the integration variables
$(\alpha_{1},T)$ to $(\varrho,\tilde{\tau})$.
For a fixed $\alpha$, 
eq.(\ref{eq:varrho}) implies that
\begin{equation}
d\alpha_1=\alpha d\varrho~, 
\qquad 
dT =
\frac{\partial T}{\partial \tilde{\tau}} d\tilde{\tau}
= \frac{\partial T}{\partial \tau} \frac{1}{\tilde{\tau}^{2}}
  d\tilde{\tau}~.
\end{equation}
We find that $\partial T / \partial \tau$
becomes
\begin{equation}
\frac{\partial T}{\partial \tau} = -\frac{i}{4\pi} c_{I}~,
\end{equation}
where
$c_{I}$ is given in eq.(\ref{eq:cI}).
This can be derived from
eqs.(\ref{eq:Mandelstam-annulus}),
(\ref{eq:int-point}) and (\ref{eq:cI})
as follows:
\begin{eqnarray}
\frac{\partial T}{\partial \tau}
 &=& \frac{\partial \rho}{\partial \nu} (\nu_{I}^{-})
       \, \frac{\partial \nu_{I}^{-}}{\partial \tau}
      + \left.
         \alpha \frac{\partial}{\partial \tau}
         \ln \frac{\vartheta_{1}(\nu + V_{3} |\tau)}
                  {\vartheta_{1}(\nu - V_{3}|\tau)}
       \right|_{\nu=\nu_{I}^{-}}
     = \left.
         \alpha \frac{\partial}{\partial \tau}
         \ln \frac{\vartheta_{1}(\nu + V_{3} |\tau)}
                  {\vartheta_{1}(\nu - V_{3}|\tau)}
       \right|_{\nu=\nu_{I}^{-}}
\nonumber\\
&=& -\frac{i}{4\pi}\alpha
    \left[
      \frac{\partial^{2}_{\nu}
            \vartheta_{1} (\nu^{-}_{I} + V_{3}|\tau)}
           {\vartheta_{1} (\nu^{-}_{I} + V_{3}|\tau)}
      - \frac{\partial^{2}_{\nu}
            \vartheta_{1} (\nu^{-}_{I} - V_{3}|\tau)}
           {\vartheta_{1} (\nu^{-}_{I} - V_{3}|\tau)}
     \right]
 = -\frac{i}{4\pi} c_{I}~.
\label{eq:dTdtau}
\end{eqnarray}
Here we have used the fact that the theta function
$\vartheta_{1} (\nu|\tau)$ satisfies
the heat equation,
\begin{equation}
\frac{\partial}{\partial \tau} \vartheta_{1} (\nu|\tau)
  = -\frac{i}{4\pi} \frac{\partial^{2}}{\partial \nu^{2}}
     \vartheta_{1} (\nu|\tau)~.
\label{eq:heateq}
\end{equation}

\subsubsection*{S-matrix element}
Collecting all these results, we can obtain
\begin{eqnarray}
\lefteqn{
G_{\phi DD'}(k) = - 4 \pi^2  gA^{2}
  \int^{\infty}_{0} d\alpha 
  \, \frac{t}{\alpha^{2}}
  \,
   e^{- \frac{t}{\alpha} (k^2 + M^2)}
  \int^{1}_{0} d\varrho
   \int^{\tilde{\tau}_{0}(\alpha,\varrho)}_{0} 
   d\tilde{\tau} \, \tilde{\tau}
 \left| 
   \frac{e^{-i\pi\varrho^{2}\tilde{\tau}}
                  \eta(\tilde{\tau})^3}
        {\vartheta _1 (\varrho\tilde{\tau} |\tilde{\tau})} 
 \right|^{- (k^2 + M^2)} 
}  
\nonumber\\
&& \ \quad
\times 
\prod_{n=1}^{\infty} 
        \left(1-e^{2\pi i n\tilde{\tau}}\right)^2
{}_X\langle B_0 (Y')|  
e^{i\pi\tilde{\tau} \varrho (L_0^X + \tilde{L}_0^X -2)}
\, V_\phi \,
e^{i\pi\tilde{\tau}(1-\varrho)  (L_0^X + \tilde{L}_0^X -2)}
|B_0(Y) \rangle_X~,~~~
\end{eqnarray}
where $\tilde{\tau}_{0}(\alpha,\varrho)$ is the value of
$\tilde{\tau}(=\tilde{\tau}(T,\alpha,\varrho ))$ when $T=t$:
$\tilde{\tau}_{0}(\alpha,\varrho) = \tilde{\tau}(t,\alpha,\varrho)$.

In order to obtain the S-matrix element $S_{\phi DD'}$ 
for the process
we are considering, we need look for the singular behavior of
$G_{\phi DD'}(k)$ near the mass-shell
of the external state, namely $k^{2} + M^2 \sim 0$.
As explained in \cite{Baba:2007je}, such singularity comes from
the region $\alpha \sim 0$ in the integration over $\alpha$,
and we find
\begin{eqnarray}
G_{\phi DD'}(k) &\sim& 
  - 4 \pi^2 gA^{2} \frac{1}{k^{2} + M^2}
  \int^{1}_{0} d\varrho 
  \int^{i\infty}_{0} 
        d\tilde{\tau} \,  \tilde{\tau} 
    \, \prod_{n=1}^{\infty} 
         \left(1-e^{2\pi i n\tilde{\tau}} \right)^2
\nonumber\\
&& 
\ \times 
{}_X\langle B_0 (Y')|  
e^{i\pi\tilde{\tau}\varrho(L_0^X + \tilde{L}_0^X -2)}
\, V_\phi \,
e^{i\pi\tilde{\tau}(1-\varrho) (L_0^X + \tilde{L}_0^X -2)}
|B_0(Y) \rangle_X~.
\end{eqnarray}
Here we have used the relation
\begin{equation}
\lim_{\alpha \rightarrow 0} \tilde{\tau}_{0}(\alpha, \varrho)
 = i \infty~.
\end{equation}
Thus we obtain the S-matrix element $S_{\phi DD'}$,
\begin{eqnarray}
S_{\phi DD'}(k) 
&=& 
  - 4 \pi^2 i gA^{2} 
  \int^{1}_{0} d\varrho 
  \int^{i \infty }_{0} 
        d\tilde{\tau} \, \tilde{\tau} 
    \, \prod_{n=1}^{\infty} 
         \left(1-e^{2\pi i n\tilde{\tau} } \right)^2
\nonumber\\
&& 
\ \times 
{}_X\langle B_0 (Y')|  
e^{i\pi\tilde{\tau} \varrho(L_0^X + \tilde{L}_0^X -2)}
\,  V_\phi \,
e^{i\pi\tilde{\tau} (1-\varrho) (L_0^X + \tilde{L}_0^X -2)}
|B_0(Y) \rangle_X
~,
\label{eq:Stdd-1}
\end{eqnarray}
where the momentum $k_{\hat{\mu}}$ $(\hat{\mu}=0,\ldots,25)$
of the vertex operator $V_{\phi}$
is subject to the on-shell condition:
$k^{2}+M^2 = 0$.
Here we have performed the Wick rotation to make 
the space-time signature Lorentzian.

In this form,
it is obvious that $S_{\phi DD'}$ is proportional
to the S-matrix element in first quantized string theory.
The factor 
$\prod_{n=1}^{\infty} \left(1-e^{2\pi i n\tilde{\tau} } \right)^2$ 
coincides with the ghost contribution to the partition function. 
As is described in Fig.~\ref{fig:annulus} (b),
the worldsheet of the process we are considering
is a one-punctured cylinder.
In eq.(\ref{eq:Stdd-1}), the S-matrix element $S_{\phi DD'}$
is expressed as an integral over the moduli space
of the one-punctured cylinder with the correct 
integration measure $\tilde{\tau}d\tilde{\tau}d\varrho$.
We notice that in this integral
the moduli space is covered completely and only once.

\subsection{Factorization of S-matrix element}
\begin{figure}[htb]
\begin{center}
\includegraphics[height=12em]{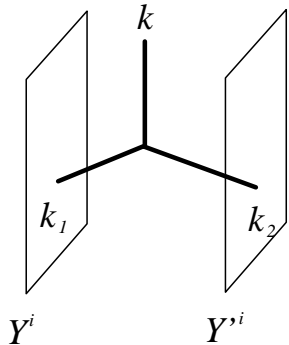}
\end{center}
\caption{The scattering process near the poles from the tachyons
         of the closed strings exchanged between the two D-branes
          in Fig.~\ref{fig:annulus} (a).
         The three solid lines connecting
         with each other indicate tachyon propagations.}
\label{fig:poles}
\end{figure}

Let us check that the S-matrix element in eq.(\ref{eq:Stdd-1})
has the correct normalization.
This can be done by considering the S-matrix element $S_{\phi DD'}$
in the simplest case
where $V_\phi$ corresponds to the tachyon:
\begin{equation}
V_\phi
 = \maru \, e^{ik\cdot X} \, \maru~.
\end{equation}
Here $\maru\ \maru$ denotes the normal ordering of the oscillators.
We examine the behavior of $S_{\phi DD'}$ at the poles from
the tachyons of the closed strings exchanged between the two D-branes
in Fig.~\ref{fig:annulus} (a).
This corresponds to the scattering process sketched in 
Fig.~\ref{fig:poles}. 
In order to obtain the singular behaviour, 
we perform the Fourier transformation of the S-matrix
element $S_{\phi DD'}$  with respect to $Y^{i}$ 
and $Y^{\prime i}$, and then put the conjugate momenta 
$k_{1}$ and $k_{2}$ close to the mass-shell of the tachyon.
In the region where 
$k_{1}^{2} \sim 2$ and $k_{2}^{2} \sim 2$,
$S_{\phi DD'}$ becomes
\begin{equation}
S_{\phi DD'} \sim
\int \frac{d^{26}k_{1}}{(2\pi)^{26}}
 \int \frac{d^{26}k_{2}}{(2\pi)^{26}}
 S_{TD}(- k_{1};Y) \frac{-i}{k_{1}^{2}-2}
 S_{TTT}(k, k_{1}, k_{2})
 \frac{-i}{k_{2}^{2} -2} S_{TD} (- k_{2};Y')~,
\label{eq:factorization}
\end{equation}
where
\begin{eqnarray}
S_{TTT} (k, k_{1}, k_{2})
 &=& i 4g (2\pi)^{26} \delta^{26} (k+k_{1}+k_{2})~,
 \nonumber\\
S_{TD}(k_{1};Y)
&=& i A (2\pi)^{p+1} \delta^{p+1}_{\mathrm{N}} (k_{1})
    e^{ik_{1,i}Y^{i}}~.
\label{eq:Smatrices}
\end{eqnarray}
In this equation,
$S_{TTT}(k, k_{1}, k_{2})$
is the tree amplitude for three closed string tachyons
with momenta $k$, $k_{1}$
and $k_{2}$, and
$S_{TD}(k_{1};Y)$ is the coupling of
the D$p$-brane located at $X^{i} = Y^{i}$ to a closed
string tachyon with momentum $k_{1}$.\footnote{
In \cite{Baba:2007tc},
we showed that $S_{TD}$ can be reproduced by
the states (\ref{eq:1Dbrane}) with one D-brane.}
Eq.(\ref{eq:factorization}), therefore, implies that
the factorization occurs in the right way in $S_{\phi DD'}$
and thus $S_{\phi DD'}$ has the correct normalization.

\subsection{More general amplitudes}
It is easy to generalize the calculation above and 
consider more general annulus amplitudes. 
For example, let us consider
the amplitudes with the annulus ending on the same D-brane. 
In this case the computations are the same as those
in the case of two D-branes, except that this time
the S-matrix elements are deduced from the contributions of
the term quadratic in the boundary state
contained only in a single $\bar{\mathcal{O}}_{D}$.
Therefore the normalizations of the S-matrix elements become
half of those in the case of two D-branes.
Thus we obtain the correct normalizations.

We can also calculate the annulus amplitudes 
with any number of closed string insertions. 
We can compute such amplitudes
by using the fact that the three-string interaction vertex 
overlapped by an external state 
reduces to the vertex operator for the state,
when the external state is close to the mass-shell
\cite{Baba:2007tc}.
Therefore the computation comes down to the one we have done 
above.
It is easy to check that 
the resulting S-matrix elements
are expressed as an integral over the moduli space
with the appropriate measure
and have 
the correct normalizations.

\section{Conclusions}
\label{sec:discussions}

In this paper, we construct states corresponding to
$N$ parallel D$p$-branes located separately from each other.
We show that these states are BRST invariant in the leading
order in $\epsilon$.
Using these states, we can calculate annulus amplitudes.
We show that usual annulus amplitudes are reproduced.
The analyses in this paper provide another evidence for 
our construction of the D-brane states
in the $OSp$ invariant closed string field theory.


\section*{Acknowledgements}

We would like to thank I.~Kishimoto for discussions.
This work was supported in part by 
Grant-in-Aid for Scientific Research (C) (20540247), 
Grant-in-Aid for Young Scientists (B) (19740164) from
the Ministry of Education, Culture, Sports, Science and
Technology (MEXT), and Grant-in-Aid for
JSPS Fellows (19$\cdot$1665).

\appendix

\section{Details of Calculation of
   $\langle V_{1} (3);Y,Y';T|$}
\label{sec:detailcalculation}

In this appendix, we present details of the calculation
of the string vertex (\ref{eq:V1YYT-1}).

\begin{figure}[htb]
\begin{center}
\includegraphics[height=14em]{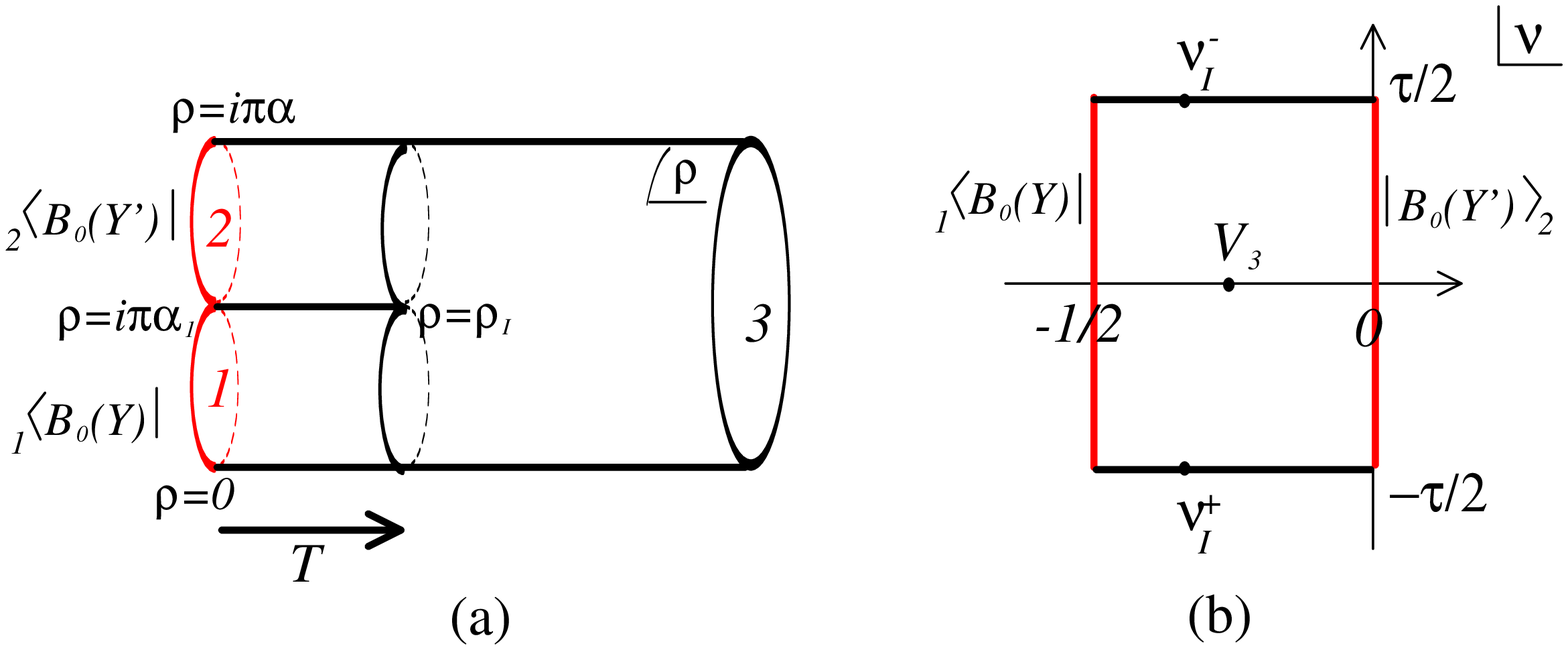}
\caption{(a) The worldsheet for the string diagram corresponding
         to the vertex (\ref{eq:V10YYT}).
         The coordinate $\rho$
           ($\mathrm{Re}\, \rho \geq 0$,
           $-\pi\alpha \leq \mathrm{Im}\,\rho \leq \pi\alpha$)
         is introduced on this worldsheet.
         (b) The rectangle on the $\nu$-plane related to the
         worldsheet in (a) by the Mandelstam mapping
         (\ref{eq:Mandelstam-annulus}).}
\label{fig:worldsheet}
\end{center}
\end{figure}

The vertex (\ref{eq:V1YYT-1}) is expressed as
\begin{equation}
\langle V_{1} (3);Y,Y';T|
 = \langle V_{1}^{0} (3); Y,Y';T| C(\rho_{I})
   \mathcal{P}_{3}~,
\label{eq:V1YYT-3}
\end{equation}
where
\begin{equation}
\langle V^{0}_{1}(3);Y,Y';T|
 \equiv \int d'1 \, d'2 \, \delta(1,2,3)
  \frac{|\mu (1,2,3)|^{2}}{\alpha_{1} \alpha_{2} \alpha_{3}}
  {}_{123} \langle 0| e^{E(1,2,3)}
  |B_{0}(Y)\rangle^{T}_{1}
  |B_{0}(Y')\rangle^{T}_{2}~.
\label{eq:V10YYT}
\end{equation}
As carried out in \cite{Baba:2007tc},
we introduce the complex coordinate $\rho$ on the worldsheet
for the string diagram corresponding to
the vertex (\ref{eq:V10YYT}) depicted in 
Fig.~\ref{fig:worldsheet} (a).
$\rho_{I}$ in eq.(\ref{eq:V1YYT-3}) denotes the interaction
point on the $\rho$-plane.
The external closed string corresponds to the string $3$. 
The region of the worldsheet corresponding to the propagation of 
this string is 
$|w_{3}|\leq 1$, where
\begin{equation}
\rho = \alpha_{3} \ln w_{3} + T~.
\label{eq:unitdisk}
\end{equation}

The vertex
$\langle V_{1}^{0}(3);Y,Y';T|$ takes the form
\begin{eqnarray}
\langle V^{0}_{1}(3);Y,Y';T|
&=& 2 \delta(\alpha_{1} + \alpha_{2} + \alpha_{3})
   (2\pi)^{p+1} \delta^{p+1}_{\mathrm{N}} (p_{3})
\nonumber\\
&&\quad  \times
    \mathcal{K}_{1}(3;Y,Y';T)
    \langle V^{0}_{1,\mathrm{LPP}} (3); Y,Y';T|~,
\label{eq:LPPrectangle}
\end{eqnarray}
where $\mathcal{K}_{1}(3;Y,Y';T)$ is the partition
function of the CFT on the $\rho$-plane endowed with the metric
$ds^{2} = d\rho d\bar{\rho}$ \cite{Mandelstam:1985ww}.
$\langle V^{0}_{1,\mathrm{LPP}} (3); Y,Y';T|$
is the LPP vertex \cite{LeClair:1988sp} of the form
\begin{equation}
\langle V^{0}_{1,\mathrm{LPP}} (3); Y,Y';T|
=
{}_3\langle 0|e^{E(3)}~,
\label{eq:E3}
\end{equation}
where $E(3)$ consists of terms
linear or quadratic in $\alpha_n^{M(3)}$ and
$\tilde{\alpha}_{n}^{M(3)}$  ($n\geq 0$).

\subsubsection*{Mandelstam mapping}

In order to evaluate the string vertex (\ref{eq:V1YYT-1}),
we use the Mandelstam mapping
introduced in \cite{Baba:2007tc},
\begin{equation}
\rho (\nu) = \alpha \ln
  \frac{\vartheta_{1} (\nu + V_{3}|\tau)}
       {\vartheta_{1} (\nu - V_{3}|\tau)}~,
\label{eq:Mandelstam-annulus}
\end{equation}
where $\alpha \equiv \alpha_{1} + \alpha_{2} = - \alpha_{3}$,
$V_{3} = -\frac{\alpha_{1}}{2\alpha}$ and
$\vartheta_{1} (\nu | \tau)$ 
is a Jacobi theta function.
This is the mapping between
the $\rho$-plane and the rectangle on the complex
$\nu$-plane defined by
$-\frac{1}{2} \leq \mathrm{Re}\, \nu \leq 0$ and
$-\frac{\tau_{2}}{2}
   \leq \mathrm{Im} \, \nu \leq \frac{\tau_{2}}{2}$
(Fig.~\ref{fig:worldsheet} (b)).
Here $\tau=i\tau_{2}$
($\tau_{2} \in \mathbb{R}$, $\tau_{2}\geq 0$)
is the modulus of the annulus and the identification
$\nu \cong \nu + \tau$ should be made.
The interaction points $\nu_{I}^{\pm}$ on the $\nu$-plane
and the modulus $T$ on the $\rho$-plane
satisfy
\begin{equation}
\frac{d\rho}{d\nu} (\nu_{I}^{\pm}) =0~,
\quad
T = \mathrm{Re} \rho(\nu_{I}^{-})
     =\rho(\nu_{I}^{-}) + 2\pi i \alpha V_{3}~.
\label{eq:int-point}
\end{equation}

\subsubsection*{Partition function $\mathcal{K}_{1}(3;Y,Y';T)$}

{}From the Mandelstam mapping (\ref{eq:Mandelstam-annulus}),
one can find that the boundary conditions imposed on
the worldsheet variables $X^{i}(\nu,\bar{\nu})$
$(i=p+1,\ldots, 25)$ on the $\nu$-plane are
\begin{equation}
\left. X^{i}(\nu,\bar{\nu})
\right|_{\mathrm{Re}\, \nu=-\frac{1}{2}}
 = Y^{i}~,
\quad
\left. X^{i}(\nu,\bar{\nu})
\right|_{\mathrm{Re} \, \nu=0}
 = Y^{\prime i}~,
\quad
X^{i}(\nu+\tau,\bar{\nu}+\bar{\tau}) = X^{i} (\nu,\bar{\nu})~,
\end{equation}
and the other worldsheet variables
$X^{\mu}(\nu,\bar{\nu})$ $(\mu=26,1,\ldots,p)$,
$C(\nu,\bar{\nu})$ and $\bar{C}(\nu,\bar{\nu})$
obey the same boundary conditions
as those in the case considered in \cite{Baba:2007tc}.
Therefore, the classical configurations
$X_{\mathrm{cl}}^{N}(\nu,\bar{\nu})$ for the worldsheet variables
around which the quantum fluctuations
$\tilde{X}^{N}(\nu,\bar{\nu})$ should be considered are
\begin{equation}
X^{i}_{\mathrm{cl}}(\nu,\bar{\nu}) 
 = Y^{\prime i} - (\nu + \bar{\nu}) \Delta Y^{i}~,
\quad
X^{\mu}_{\mathrm{cl}} (\nu,\bar{\nu}) = 0~,
\quad
C_{\mathrm{cl}}(\nu,\bar{\nu})
  = \bar{C}_{\mathrm{cl}}(\nu,\bar{\nu})=0~.
\label{eq:classicalXi}
\end{equation}

Dividing $X^{N}(\nu,\bar{\nu})$ as
$X^{N}(\nu,\bar{\nu}) = X_{\mathrm{cl}}^{N} (\nu,\bar{\nu})
  + \tilde{X}^{N}(\nu,\bar{\nu})$,
we compute the annulus partition function
$Z(\tau,\Delta Y)$ on the $\nu$-plane
(other than the effects of the puncture $\nu = V_{3}$ and
 the interaction points $\nu=\nu_{I}^{\pm}$).
We find that
\begin{equation}
Z(\tau,\Delta Y) = \int [dX]  \,  e^{-S[X]}
=e^{-S_{\mathrm{cl}}} \tilde{Z}(\tau)~,
\label{eq:partitionfnc}
\end{equation}
where $S[X]$ is the worldsheet action and
$S_{\mathrm{cl}}$ denotes its classical value
given by
\begin{eqnarray}
S[X] &\equiv& \frac{1}{2\pi}
   \int^{0}_{-\frac{1}{2}} d(\mathrm{Re}\, \nu)
   \int^{\frac{\tau_{2}}{2}}_{-\frac{\tau_{2}}{2}}
       d(\mathrm{Im}\, \nu)  \,
    \partial_{\nu} X^{N} \partial_{\bar{\nu}} X^{M} \eta_{NM}~,
 \nonumber\\
S_{\mathrm{cl}} &\equiv& S[X_{\mathrm{cl}}]
= -i \frac{\tau}{4\pi} \left( \Delta Y^{i} \right)^{2}~,
\label{eq:classicalaction}
\end{eqnarray}
and $\tilde{Z}(\tau)$ is the contribution of the fluctuations
to the partition function.
One can find that $\tilde{Z}(\tau)$ equals to the partition
function in the case where $Y^{i}=Y^{\prime i}=0$.
Combined with eq.(\ref{eq:partitionfnc}), this implies
that
\begin{equation}
\mathcal{K}_{1} (3; Y,Y';T)
 = e^{-S_{\mathrm{cl}}}
   \mathcal{K}_{1} (3;T)
 = e^{-S_{\mathrm{cl}}}
\frac{(2\pi)^{p+1}}{(2\pi)^{23}}
 \frac{e^{\frac{2T}{\alpha}}}
      {(-i\tau)^{\frac{p+1}{2}} \eta(\tau)^{18}
       \alpha^{2} \, c_{I} \,
       \vartheta_{1}  (2V_3|\tau)^{2}}~,
\label{eq:partition-annulus}
\end{equation}
where
$\mathcal{K}_{1} (3;T)=\mathcal{K}_{1}(3;0,0;T)$
is evaluated in \cite{Baba:2007tc} and
$c_{I}$ is
\begin{equation}
c_{I} \equiv \frac{d^{2}\rho}{d\nu^{2}} (\nu_{I}^{-})
  = \alpha
   \left(
      \frac{\partial^{2}_{\nu}
            \vartheta_{1}(\nu_{I}^{-} + V_{3}|\tau)}
           {\vartheta_{1}(\nu_{I}^{-} + V_{3}|\tau)}
      - \frac{\partial^{2}_{\nu}
                 \vartheta_{1}(\nu_{I}^{-} - V_{3}|\tau)}
             {\vartheta_{1} (\nu_{I}^{-} - V_{3}|\tau)}
   \right)~.
\label{eq:cI}
\end{equation}

\subsubsection*{LPP vertex
$\langle V_{1,\mathrm{LPP}}^{0}(3);Y,Y';T|$}

The LPP vertex
$\langle V_{1,\mathrm{LPP}}^{0}(3);Y,Y';T|$
introduced in eq.(\ref{eq:E3})
can be determined by the equations
\begin{eqnarray}
&& \int d'3 \, \langle V^{0}_{1,\mathrm{LPP}} (3);Y,Y';T|
     X^{N(3)}(w_{3},\bar{w}_{3}) |0\rangle_{3}
     (2\pi)^{26} \delta^{26} (p_{3})
     i\bar{\pi}_{0}^{(3)} \pi_{0}^{(3)}
\nonumber\\
&& \qquad \quad 
   =\frac{\langle X^{N}(\nu_{3},\bar{\nu}_{3}) \rangle}
         {Z(\tau,\Delta Y)}
\equiv \frac{1}{Z(\tau,\Delta Y)}
     \int [dX] \, X^{N}(\nu_{3},\bar{\nu}_{3}) \, e^{-S[X]}
 = X^{N}_{\mathrm{cl}} (\nu_{3},\bar{\nu}_{3})~,
\nonumber\\
&& \int d'3 \, \langle V^{0}_{1,\mathrm{LPP}} (3);Y,Y';T|
    X^{N (3)}(w_{3},\bar{w}_{3}) 
    X^{M (3)}(w'_{3},\bar{w}'_{3})
   |0\rangle_{3}
     (2\pi)^{26} \delta^{26} (p_{3})
     i\bar{\pi}_{0}^{(3)} \pi_{0}^{(3)}
\nonumber\\
&& \qquad \quad
  = \frac{\langle X^{N}(\nu_{3},\bar{\nu}_{3})
           X^{M}(\nu'_{3},\bar{\nu}'_{3})\rangle}
         {Z(\tau,\Delta Y)}
\equiv \frac{1}{Z(\tau,\Delta Y)}
   \int [dX] \, X^{N}(\nu_3,\bar{\nu}_3) X^{M}(\nu'_3,\bar{\nu}'_3)
   \, e^{- S[X]}
\nonumber\\
  && \qquad \quad
   = X^{N}_{\mathrm{cl}} (\nu_3,\bar{\nu}_3)
      X^{M}_{\mathrm{cl}} (\nu'_3,\bar{\nu}'_3)
    + G^{NM}_{\mathrm{rectan.}}
         (\nu_3,\bar{\nu}_3;\nu'_3,\bar{\nu}'_3)~,
\label{eq:annulus-correlation-lpp}
\end{eqnarray}
where $\nu_{3}$ and $\nu'_{3}$ are the points on the $\nu$-plane
corresponding to the points $w_{3}$ and $w'_{3}$ $(|w_3|,|w_3^\prime |<1)$,
and
$G^{NM}_{\mathrm{rectan.}} (\nu,\bar{\nu};\nu',\bar{\nu}')$
are the two-point functions of $X^{N}(\nu,\bar{\nu})$
given in \cite{Baba:2007tc}
in the case of $Y^{i} = Y^{\prime i} =0$.
This yields
\begin{equation}
\langle V_{1,\mathrm{LPP}}^{0}(3);Y,Y';T|
 = \langle V_{1,\mathrm{LPP}}^{0}(3);T|
   e^{i\sum_{n=0}^{\infty}(\bar{N}^{h}_{n,i}\alpha^{i(3)}_{n}
          + \bar{N}^{a}_{n,i} \tilde{\alpha}^{i(3)}_{n})}~,
\end{equation}
where 
$\langle V_{1,\mathrm{LPP}}^{0}(3);T|
  = \langle V_{1,\mathrm{LPP}}^{0}(3);0,0;T|$
is the LPP vertex computed in \cite{Baba:2007tc},
and the Neumann coefficients $\bar{N}^{h}_{n,i}$
and $\bar{N}^{a}_{n,i}$ are
\begin{eqnarray}
\bar{N}^{h}_{n,i}
= \left( \bar{N}^{a}_{n,i} \right)^{\ast}
&=&\frac{1}{n} \oint_{V_{3}} \frac{d\nu}{2\pi i}
  \left(w_3(\nu )\right)^{-n} \partial_{\nu} X_{\mathrm{cl}}^{i} (\nu)
=-\frac{\Delta Y^{i}}{n}\oint_{V_{3}} \frac{d\nu}{2\pi i}
  \left(w_3(\nu )\right)^{-n}
\quad \mbox{for $n\geq 1$}~,
\nonumber\\
\bar{N}^{h}_{0,i} + \bar{N}^{a}_{0,i}
  &=& X^{i}_{\mathrm{cl}}(V_{3},V_{3})
  = -2 V_{3} Y^{i} + (1 + 2 V_{3}) Y^{\prime i}~.
\label{eq:neumann-1pt-annulus}
\end{eqnarray}

Collecting all the results obtained in the above, we have
\begin{equation}
\langle V_{1}^{0}(3);Y,Y';T|
=e^{-S_{\mathrm{cl}}}
 \langle V_{1}^{0}(3);T|
   e^{i\sum_{n=0}^{\infty}(\bar{N}^{h}_{n,i}\alpha^{i(3)}_{n}
          + \bar{N}^{a}_{n,i} \tilde{\alpha}^{i(3)}_{n})}~,
\label{eq:V10YYT-2}
\end{equation}
where $\langle V_{1}^{0}(3);T| = \langle V_{1}^{0}(3);0,0;T|$
is evaluated in \cite{Baba:2007tc}.

\subsubsection*{Ghost field insertion}

Finally, we consider the effect of the insertion of the
ghost field in the vertex
$\langle V_{1}^{0}(3);Y,Y';T|$
to obtain $\langle V_{1}(3);Y,Y';T|$.
This is the same
as that obtained in \cite{Baba:2007tc}.
Eventually, we obtain 
\begin{eqnarray}
\lefteqn{
\langle V_{1}(3);Y,Y';T|
}  \nonumber\\
&& = 
 e^{-S_{\mathrm{cl}}}
   \langle V_{1}^{0} (3);T|
    i \sum_{n=0}^{\infty} \left(
   {M_{\mathrm{rectan.}}}^{h}_{n} \gamma^{(3)}_{n}
   + {M_{\mathrm{rectan.}}}^{a}_{n} \tilde{\gamma}^{(3)}_{n}
   \right) 
   e^{i\sum_{n=0}^{\infty} 
       \left( \bar{N}^{h}_{n,i} \alpha^{i(3)}_{n}
              + \bar{N}^{a}_{n,i} \tilde{\alpha}^{i(3)}_{n}
       \right)}
   \mathcal{P}_{3}~.~~~~~
\label{eq:V1YYT-2}
\end{eqnarray}

\subsubsection*{Limit of $T=\epsilon \rightarrow 0$}

In the $T=\epsilon \rightarrow 0$ limit,
$\bar{N}^{h}_{n,i}$ and $\bar{N}^{a}_{n,i}$
for $n \geq 1$ become
\begin{equation}
\bar{N}^{h}_{n,i},
\  \bar{N}^{a}_{n,i}
\simeq i \frac{\Delta Y^{i}}{n}
     \frac{e^{-n\frac{\epsilon}{\alpha}}}{2\pi}
     \left(e^{in2\pi V_{3}} - e^{-in 2\pi V_{3}} \right)
     \left(1+ \mathcal{O}(\epsilon^2)\right)~,
\label{eq:neumann-epsilon}
\end{equation}
and thus finite.
Combined with eq.(\ref{eq:sclepsilon}),
this yields the suppression stated in 
section \ref{sec:ND-braneState}, and 
one can deduce that the states (\ref{eq:NDbranes}) 
are BRST invariant in the leading order in $\epsilon$.

In this limit,
eq.(\ref{eq:V1YYT-2}) becomes
the idempotency equation \cite{Kishimoto:2003ru}
in the $OSp$ invariant string field theory~\cite{Baba:2007tc}.
By taking the limit
$
\epsilon^{\frac{(\Delta Y^{i})^{2}}{4\pi^{2}}}
  \sim 
  \left( \frac{4\pi^{3}}{-\ln \epsilon} 
  \right)^{\frac{25-p}{2}}
  \delta^{25-p}(\Delta Y)
$
first in eq.(\ref{eq:sclepsilon}),
one can find that eq.(\ref{eq:V1YYT-2}) turns out to take
a form similar to that given in \cite{Kishimoto:2003ru}.


\end{document}